\documentclass[aps,prb,reprint,groupedaddress]{revtex4-1}

\usepackage[dvips]{graphicx}
\usepackage{amsbsy}
\usepackage{amssymb}      
\usepackage{amsmath}

\begin{document}

\title{First-principles analysis of the intermediate band in CuGa$_{1-x}$Fe$_x$S$_2$}

\author{J. Koskelo}
\affiliation{Department of Physics, University of Helsinki, P.O. Box 64, FI-00014 Helsinki, Finland}

\author{J. Hashemi}
\affiliation{Department of Physics, University of Helsinki, P.O. Box 64, FI-00014 Helsinki, Finland}

\author{S. Huotari}
\affiliation{Department of Physics, University of Helsinki, P.O. Box 64, FI-00014 Helsinki, Finland}

\author{M. Hakala}
\affiliation{Department of Physics, University of Helsinki, P.O. Box 64, FI-00014 Helsinki, Finland}

\date{\today}

\begin{abstract}
  We present a comprehensive study of the electronic, magnetic, and optical properties of CuGa$_{1-x}$Fe$_x$S$_2$, as a promising candidate for intermediate-band (IB) solar cells. We use hybrid exchange-correlation functional within the density functional theory framework, and show that Fe doping induces unoccupied states 1.6-1.9 eV above the valence band. The IBs significantly enhance the optical absorption in lower energy part of the spectrum. We find that at moderate $n$-type co-doping concentration, the added charge occupies part of the IB in the gap, but large concentrations lower the energy of the occupied IB toward the valence band. Moreover, we show that Fe impurities tend to cluster within the compound and they choose antiferromagnetic ordering. The findings can have a significant effect in understanding this material and help to synthesize more efficient IB solar cells. 
\end{abstract}


\maketitle

\section{Introduction}

The production of clean and renewable energy with moderate costs is one of the biggest aims of modern technology. Photovoltaic devices hold great promise for this task, but higher efficiencys and lower costs are still needed. Intermediate-band (IB) solar cells \cite{luque1997a,luque2012a} are one of the most viable candidates among photovoltaic technologies. In IB materials, additional bands are formed between the valence and conduction bands. These induced bands act as stepping stones in the absorption process and help to utilize a wider range of the solar spectrum. In the case of a single half-filled IB the theoretical efficiency limit \cite{marti2008a} of 47 \% clearly exceeds the Shockley-Queisser limit \cite{shockley1961a} of 31 \% for single-gap solar cell under 1 sun illumination. 

One of the ways to manufacture IB solar-cell absorbers is to dope a suitable host semiconductor material with metal impurities. CuGaS$_2$ is a promising host material having a bandgap close to the optimal value for a single half-filled IB \cite{marti2008a}, and several different dopants have been studied theoretically \cite{hashemi2014a,han2014a,chen2013a,yang2013a,seminovski2011a,tablero2010a,tablero2010b,aguilera2010a,aguilera2008a,palacios2008a,palacios2007a,palacios2006a,zongyan2014a}. Moreover, at least four elements have been experimentally shown to induce IBs \cite{chen2013a,yang2013a,lv2014a,marsen2012a}. Optical properties of CuGa$_{1-x}$Fe$_x$S$_2$ have been investigated experimentally in both bulk \cite{teranishi1974a,bardeleben1978a,sato1980a,tanaka1989a}, and more recently, in thin-film \cite{marsen2012a} form. Sub-gap absorption peaks are observed  \cite{marsen2012a,tanaka1989a} around 1.2 eV and 1.9 eV. In the case of an actual photovoltaic device, low Fe concentrations ($x$=0.003) were found to be more efficient than the larger ones due to less pronounced non-radiative recombination \cite{marsen2012a}. Unfortunately, in first principles modeling one is often restricted to higher concentrations due to computational limitations with large supercells. In the present study the concentration is $x$=0.125. However, calculations with large doping concentrations are useful for obtaining more insight on the materials, and changes that occur via doping. 

Fe-doped CuGaS$_2$ has been studied before using first-principles theoretical modeling \cite{tablero2010a,han2014a,zongyan2014a}. Tablero and Fuertes Marr\'on \cite{tablero2010a} studied thermal stability, electronic structure and magnetization of CuGa$_{1-x}$Fe$_x$S$_2$ using local density approximation, and report an unoccupied IB. Furthermore, they suggested that Fe dopants prefer antiferromagnetic ordering when substituted on gallium site. However, more advanced computational methods going beyond standard density functional theory with local density or generalized gradient approximations are desired to reliably model the electronic properties, such as band gaps and positions of IBs within the band gap. Han \emph{et al.} \cite{han2014a} studied CuGa$_{1-x}$Fe$_x$S$_2$ using a hybrid exchange-correlation (XC) functional, and also found an unoccupied IB. They considered a cell with one Fe impurity in which case a ferromagnetic ordering is enforced. 

In this work, we employ a hybrid XC functional to study the electronic structure of CuGa$_{1-x}$Fe$_x$S$_2$ considering both ferro- and antiferromagnetic orderings. We substitute gallium with Fe impurities, and take into account structures with varying dopant distances. We show that the antiferromagnetic ordering and short distances between the dopants are energetically preferred. We find unoccupied IBs between 1.6 and 1.9 eV above the valence band edge. These IBs can be straightforwardly attributed to the higher energy sub-gap peak in optical absorption experiments. The lower peak around 1.2 eV observed in experiments is not as easy to explain, and it may be related to a defect structure not taken into account in our calculations. Motivated by the ideal half-filled picture of the IB solar cell, we also study the effect of charge addition to obtain filled bands in the gap. We observe that at moderate $n$-doping concentrations the added electron fills a state that remains in the gap. But, increasing the concentration lowers the energy of this state toward the top of the valence band, which implies possible constrains in obtaining more efficient IB solar cells by high concentrations of $n$-type co-doping.

\section{Computional methodology}

\subsection{Model structures and geometry optimizations}

We have performed most of the calculations in this work on rectangular 64-atom supercells including 2 Fe impurities. This allows to study both the effect of different relative distances and magnetization states. We consider substitutional doping on gallium sites, since Fe can easily acquire the +3 charge state that also gallium has. We do not take substitution on Cu sites into account since Cu has +1 charge state which is not usual for Fe. We divide all these doped structures into four classes based on the distance between the dopants where each class contains $f$ different sructures with same dopant distance, as in Ref. \citenum{hashemi2014a}.  We denote these classes as S-CGS:Fe, M-CGS:Fe, L-CGS:Fe, and XL-CGS:Fe, in the order of increasing interatomic distance between the dopants (3.8, 5.4, 6.5 and 7.6 \AA, respectively) as they are summarized in Table \ref{energytable}. In L-CGS:Fe class there are structures with two slightly different distances (6.49 and 6.54 {\AA}) and only one of these is considered. In other classes the structures are completely equivalent. 

The supercell dimensions ($a$=10.72 and $c$=10.54 \AA) were obtained from the lattice parameters of pristine CuGaS$_2$ calculated in Ref. \citenum{hashemi2014a}. To check the effect of Fe doping on the cell parameters, we relaxed a 16-atom cell with one Fe. The change in the cell volume was only 1.04 \%, and thus we used the pristine cell parameters for the larger supercells. However, we relaxed the atomic positions in all the supercells. In all the geometry optimizations in this work, we used Perdew-Burke-Ernzerhof (PBE) functional \cite{perdew1996a}, plane-wave cutoff energy of 400 eV, and 2x2x1, 2x2x2 and 4x4x2 Monkhorst-Pack ${\bf k}$-point samplings for the 128-, 64-, and 16-atom cells, respectively. To extend the conclusions from our 64-atom calculations, we performed calculations for two 128-atom structures: S-CGS:Fe duplicated in z-direction and one where the four Fe atoms were moved to a configuration with smallest possible mean distance between them. 

\subsection{Electronic structure and optical calculations}

We performed the electronic structure calculations using the hybrid XC functional according to Heyd, Scuzeria, and Ernzerhof (HSE06) \cite{heyd2003a,heyd2006a} that mixes 25 \% of short-range exact exchange with the exchange of the PBE functional. HSE06 is widely used in modeling electronic properties of Cu-based chalcopyrite materials and gives a reasonable description of band gap in pristine CuGaS$_2$ (2.18 eV vs. experimental range \cite{shay1972a,tell1971a,bellabarba1996a,syrbu2005a,botha2007a} of 2.4-2.53 eV). We employed projector augmented wave (PAW) \cite{blochl1994a} method and VASP \cite{kresse1996a} simulation package for all the calculations in this work. For the 64-atom supercell we used 3$\times$3$\times$3 $\Gamma$-centered ${\bf k}$-point sampling (the 128-atom calculation was performed with 3$\times$3$\times$1). In the construction of the Fock operator we downsampled the ${\bf k}$-grid with a factor of 3 in each direction. We used 400 eV as the plane-wave cutoff energy. We performed the calculations as spin dependent and forced the 5 $\mu_B$ Fe magnetic moments to be either in ferromagnetic or antiferromagnetic configurations. We also tested the influence of ${\bf k}$-point grid up to 7$\times$7$\times$7 (with reduction of 7 for exact exchange) on the density of states, and the IBs were practically unaffected by that, the main effect being on the valence and conduction bands (see supplementary material). 

The optical spectra were calculated within the random phase approximation (RPA) using the bands from hybrid-functional calculations. Due to the highly dispersed first conduction band, the calculations of optical spectra require dense ${\bf k}$-point sampling. For 64-atom cells the sampling was 7$\times$7$\times$7 and for 16-atom pristine cell 14$\times$14$\times$7, and for the exact exchange it was reduced by a factor of 7 in each direction. To evaluate the convergence with respect to ${\bf k}$-points, the optical spectra using different ${\bf k}$-point samplings in pristine cell is presented in supplementary material. Local fields are taken into account by considering ${\bf G}$-vectors up to 50 eV in the inversion of the dielectric function. The Lorentzian broadening applied for the optical spectra was 0.1 eV.

\section{Results and discussion}

\subsection{Total energies}

To evaluate the probabilities of different structural and magnetic configurations in the thermal equilibrium, we report the total energies of the four aforementioned structures in Table \ref{energytable} both in ferro- and antiferromagnetic orderings. The statistical weights for the configurations in the room temperature are evaluated using the Boltzmann distribution, following Ref. \citenum{hashemi2014a}. The results show that antiferromagnetic ordering is in overall preferred for all considered structures. The energetically most preferred configuration is the antiferromagnetic S-CGS:Fe, \emph{i.e.} the structure with shortest distance between the dopants. Not surprisingly, the energy difference of ferromagnetic and antiferromagnetic ordering is largest with S-CGS:Fe, which means that the magnetic interaction is largest at smallest distance. However, this energy difference does not decrease monotonously with increasing distance, since the magnetic interaction can be mediated in a non-trivial fashion as a function of distance. It should be noted, that the energy differences are qualitatively similar when calculated using the PBE functional \cite{perdew1996a}, except that the differences are significantly larger (for example, the difference of antiferromagnetically ordered S- and XL-CGS:Fe is 100 meV with PBE compared to 46 meV with HSE06). The energetic preference for antiferromagnetic ordering corroborates a previous study using local density approximation \cite{tablero2010a}. Furthermore, the compound corresponding to extremal Fe doping, CuFeS$_2$, is known to be antiferromagnetic. 

The remaining of this article focuses mostly on the electronic structure and optical properties of antiferromagnetic S-CGS:Fe, since this has the largest statistical weight in thermal equilibrium. 

\begin{table} 
\caption{Total energy differences $\Delta E$ for different classes of doped structures along with number of structures $f$ in each class, distance $R$ between the dopants, and statistical weight $W$ of the configuration in thermal equilibrium.}
\begin{ruledtabular}
\begin{tabular}{llllllllllllllll}
configuration & \, $f$ & \, $R$ ({\AA}) & \, $\Delta E$ (meV) & \, $W$\\
\hline
S-CGS:Fe &  \, 4 &  \,  3.81 \\
$\uparrow \uparrow$ &  \, &  \,&  \,  182 &  \, 0.0007\\ 
$\uparrow \downarrow$ &  \,&  \,&  \,  0 &  \, 0.7454  \\
M-CGS:Fe &  \, 2 &  \, 5.36 \\
$\uparrow \uparrow$ &  \, &  \,&  \,    81 &  \, 0.0162 \\
$\uparrow \downarrow$ &  \, &  \,&  \,   77 &  \, 0.0191 \\
L-CGS:Fe &  \, 8 &  \, $\sim$6.5 \\
$\uparrow \uparrow$ &  \,&  \,&  \,  86 &  \, 0.0535 \\
$\uparrow \downarrow$ &  \,&  \,&  \,  63 &  \, 0.1297 \\
XL-CGS:Fe &  \, 1 &  \, 7.58 \\
$\uparrow \uparrow$ &  \, &  \,&  \, 98 &  \, 0.0042 \\
$\uparrow \downarrow$ &  \, &  \,&  \,  46 &  \, 0.0314 \\ 
\end{tabular}
\end{ruledtabular}
\label{energytable}
\end{table}

\subsection{Ground-state electronic structure}

\begin{figure*}
\includegraphics[width=\textwidth]{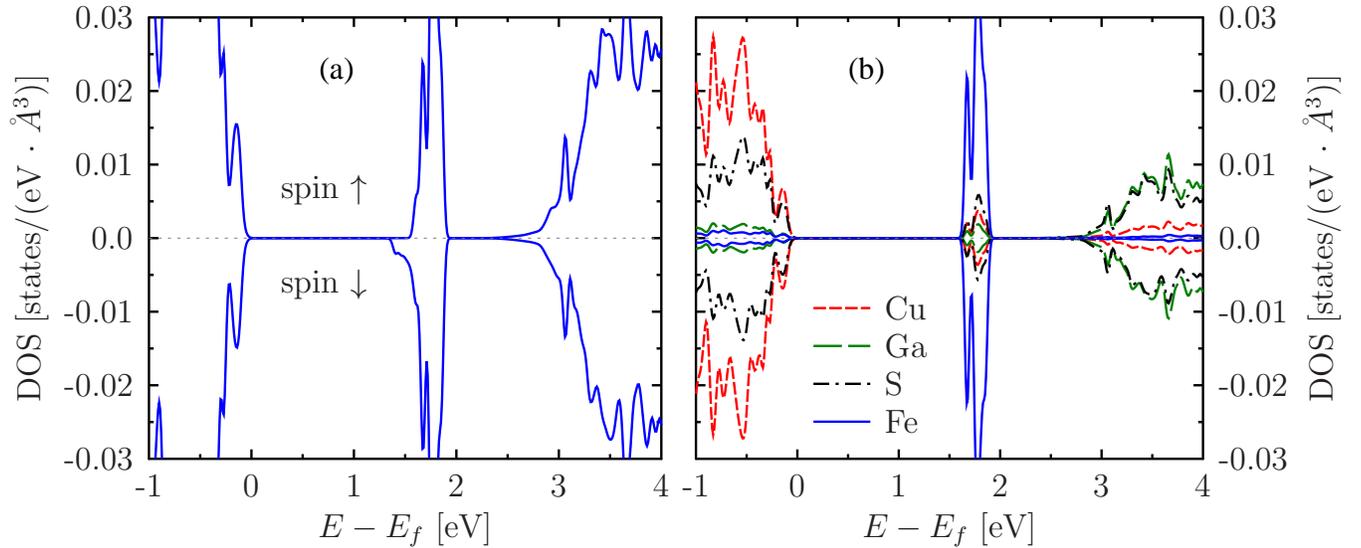}%
\caption{(a) Total density of states averaged over microstructures, and (b) atom-specific densities of states of antiferromagnetic S-CGS:Fe. All densities of states in this work are convoluted with a Gaussian of 0.05 eV full width at half maximum.}
\label{dosfig}
\end{figure*}

We present the density of states (DOS) statistically averaged over all the considered microstructures in Fig. \ref{dosfig} along with the atom-specific DOS of antiferromagnetic S-CGS:Fe. In S-CGS:Fe the unoccupied IBs are formed between 1.6 and 1.9 eV above the valence band maximum (VBM). The relative IB positions within the gap are qualitatively different when using the PBE functional. According to the PBE, there are two IBs lower in the gap and the three upper ones are merged with each other and partially overlap with the conduction band. 

The IBs mainly originate from Fe $d$ orbitals, but also have clearly visible contributions from other atoms showing the effect of hybridization. The valence band has predominantly Cu $d$ and S $p$ character, while the conduction band is mainly derived from the Ga $s$ and S $s$ and $p$ orbitals. Since the real experimentally produced structures can be affected by different conditions and sample preparations techniques, we report the total DOS for all the configurations in supplementary material. Compared to antiferromagnetic ordering, the IBs are systematically spread wider in energy in ferromagnetic configurations. 

\begin{figure}
\includegraphics[width=\columnwidth]{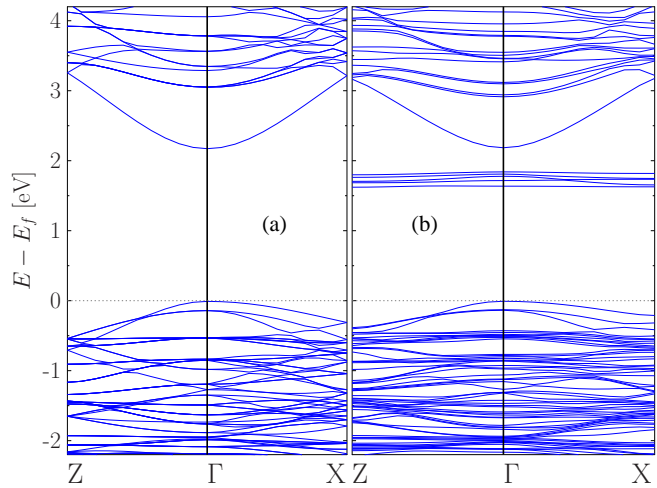}%
\caption{Band structures of (a) pristine CuGaS$_2$ and (b) S-CuGaS$_2$:Fe calculated for 64-atom supercells.}
\label{bandfig}
\end{figure}

We computed the band structures of S-CGS:Fe and pristine CuGaS$_2$ and the result is shown in Fig. \ref{bandfig}. To ease the comparison to the Fe-doped structure, the band structure of pristine CuGaS$_2$ is also given corresponding to the 64-atom supercell. The IBs are only weakly dispersed, as can be expected for localized $d$ electrons. Except for the appearance of IB states, the band structure in most parts is quite similar in the doped and pristine cases. Most notably, the dopant seems to have a negligible effect on the band gap of the host material. In the present study the gap opens by 0.01 eV, while in the previous hybrid-functional study \cite{han2014a} this effect was 0.13 eV when using twice as large dopant concentration than in our work. Also, the IBs in our work are significantly less spread in energy than in the previous work \cite{han2014a}, where the IBs were located between 1.05 eV and 1.76 eV. 

\subsection{Dopant clustering}

Above, we found that shorter Fe-Fe distances are energetically preferred for the antiferromagnetic state. The same kind of effect has previously been found in total energy calculations of Cr-doped CuGaS$_2$ \cite{palacios2008a}. To further elaborate this finding we calculated total energies for two different 128-atom structures with antiferromagnetic spin configuration and the result shows likewise the tendency of Fe dopants to cluster. The structure with smaller mean distance between the four dopants has 46 meV lower total energy than the duplicated S-CGS:Fe. To what extent clustering actually occurs depends also on the diffusion of dopants, which is beyond the scope of this work. CuGa$_{1-x}$Fe$_x$S$_2$ thin films have been experimentally observed to segregate into Ga and Fe rich phases at temperatures above 400$^{\circ}$ C. This could hint for the preference for Fe atoms to be close to each other, although the issue of phase segregation and stability is generally a complex one. 

\subsection{Optical absorption}

\begin{figure}
\includegraphics[width=\columnwidth]{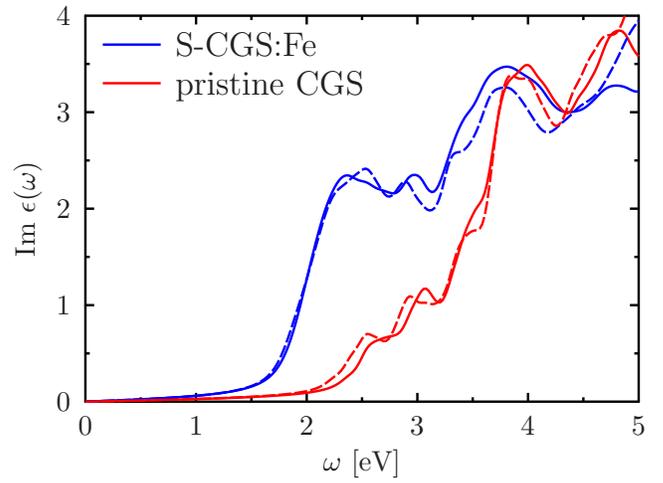}%
\caption{Optical spectra with the polarization of the light parallel to the $a$- (solid line) and $c$-axis (dashed line).}
\label{optfig}
\end{figure}

We calculated the optical absorption spectra of the antiferromagnetic S-CGS:Fe, as well the spectrum of pristine cell for comparison. The spectra of light polarized along $a$- and $c$-directions are presented in Fig. \ref{optfig}. The anisotropy between these directions is rather small, and the result for pristine cell is moreover quite similar to the GW-RPA result of Ref. \citenum{aguilera2011a} after a constant shift of the spectrum. We can see from the spectra that Fe doping clearly increases the absorption on low photon energies. 

The first peak in the absorption spectrum of S-CGS:Fe, that originates from valence-to-IB transitions, is at notably higher energy than the difference of VBM and IB. This is due to the fact that DOS has a peak around 0.5 eV below VBM. This feature is located at a higher energy than either of the experimentally observed sub-gap peaks. However, in our RPA absorption calculation excitonic effects due to electron-hole attraction are neglected, and these effects have a tendency to transfer spectral weight to lower energies. Since CuGaS$_2$ is known to have a bound exciton \cite{aguilera2011a,syrbu2005a}, and in our band-structure calculation the IBs are 1.6-1.9 eV above the VBM, excitonic effects are expected to result in a major shift of the first absorption peak to lower energy. It is thus natural to ascribe the IBs in our calculations to the 1.9 eV feature observed in experiments \cite{marsen2012a,tanaka1989a}.  

Explaining the experimental 1.2 eV peak is not as straightforward. In a previous study \cite{tanaka1989a} on the optical absorption, both sub-gap peaks were attributed to states stemming from 3$d$ orbitals of substitutional Fe. The 1.2 eV peak was assigned to a crystal field resonance state $e^{CFR}$, while the 1.9 eV peak was attributed to a dangling-bond hybrid state $t_2^{DBH}$. More advanced treatment of many-body correlation effects or detailed structure around substitutional Fe could result in larger splitting of Fe 3$d$-derived levels than in our work, and consequently to 1.2 and 1.9 eV absorption features. Another possibility, which would fit better with our results, is that the lower energy peak is due to an additional defect structure that is not yet included in this study. 

\subsection{Electron co-doping}

\begin{figure}
\includegraphics[width=\columnwidth]{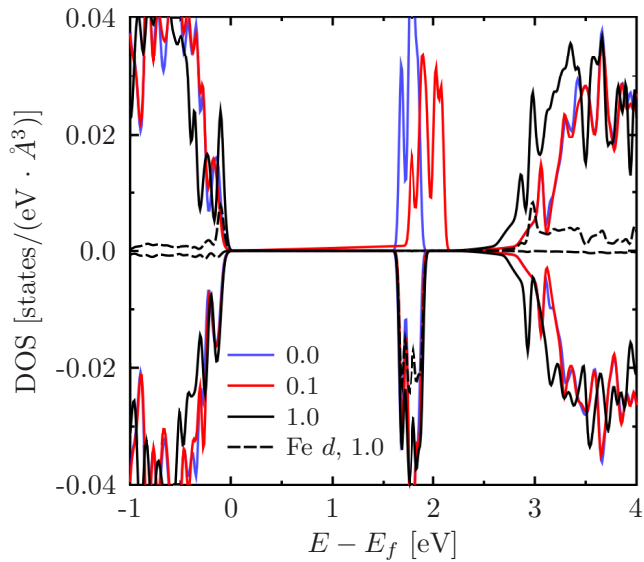}%
\caption{Density of states in S-CGS:Fe with 0.0, 0.1 and 1.0 electrons per supercell added to simulate the effect of $n$-type doping.}
\label{addelfig}
\end{figure}

Our calculations suggested that the IBs of Fe-doped CuGaS$_2$ are empty, as also reported in previous theoretical studies \cite{tablero2010a,han2014a}. However, the efficiency of the solar cell improves significantly when the IB is partially occupied, as the transitions from IB to conduction band become more probable. Therefore, it would be desirable, if we can somehow partially fill the IB, for example by $n$-type doping. Both n- and $p$-type doping of CuGaS$_2$ have been achieved upon incorporation of Zn impurities \cite{kumakura2000a}, depending on the sample preparation process. 

To simulate the effect of $n$-doping, we calculated the electronic structure with additional electronic charges keeping the total magnetic moment equal to the number of added electrons. At the smallest concentration we studied, 0.0001 electrons per supercell ($8 \times 10^{16}$ cm$^{-3}$), the added charge occupied the bottom of IB, inducing a 0.13 eV gap between occupied and unoccupied IB. Larger concentrations shift the occupied IB down, further apart from unoccupied one. When the concentration was increased by a factor of ten, the energy difference between the VBM and the occupied IB was found to become only 0.19 eV. However, the behaviour with charge addition was sensitive to $\bf k$-point sampling and smearing method, and hence these concentrations should only be considered being indicative of the order of magnitude. This behaviour may also depend on the Fe concentration. Nonetheless, the general trend from the calculations is evident, and at light $n$-type co-doping we consider the IB to be partially filled. The concentrations where we see the levels shifting to valence band are quite large, although achievable in semiconductors generally. The range of possible $n$-doping concentrations in CuGaS$_2$ is not known, but for example the extremely concentrated $n$++ region in silicon corresponds to $>$10$^{20}$ cm$^{-3}$. We note that the behaviour seen here with the HSE functional is contrasted by the one obtained using the PBE functional. In the case of PBE, the filled IB remains in the gap even at high concentrations, such as 1 electron per supercell.

In Fig. \ref{addelfig}, we report the DOS for two different electron concentrations where the occupied part of IB is moved on top of the valence band. The top of the valence band of the spin-up component then has a clear contribution from Fe $d$ orbitals. Moreover, the unoccupied IBs shift towards the conduction band with increasing additional charge, eventually coalescing with the conduction band. Meanwhile, the unoccupied Fe $d$ DOS becomes more and more spread in energy. These results serve as a demonstration of relaxations at high co-doping levels, and imply limitations in obtaining partially filled IBs in Fe-doped CuGaS$_2$. 

\section{Conclusions}

We have comprehensively studied the dopant configuration, electronic structure and optical absorption of Fe-doped CuGaS$_2$ using a hybrid exchange-correlation functional in the density functional theory framework in order to further understand the electronic structure and properties of this solar-cell material. We found that Fe dopants prefer antiferromagnetic ordering, and the dopants tend to cluster in the antiferromagnetic state. The IBs were positioned between 1.6-1.9 eV above the valence band edge. The optical spectra show enhanced absorption properties with Fe doping, although our results do not reproduce the presence of two distinct sub-gap absorption peaks observed in experiments \cite{marsen2012a,tanaka1989a}. We also studied the possible effect of an electron addition in the view of obtaining partially filled IBs, in prospect of reaching a higher solar-cell efficiency. We observed that, while typical $n$-doping concentrations of added charge result in filled band in the gap, larger concentrations lead to significant relaxations in the electronic structure so that the filled part of IB shifts to the top of the valence band. 

\begin{acknowledgments}
This work has been supported by the MATRENA Doctoral Programme, Academy of Finland (contract numbers 1256211, 1254065, 1259599, 1260204, 1259526, 1283136), and V\"ais\"al\"a foundation.
We gratefully acknowledge the generous computational resources provided by the CSC - IT Center for Science. We thank Miaomiao Han for a useful discussion. 
\end{acknowledgments}


\begin{thebibliography}{35}%
\makeatletter
\providecommand \@ifxundefined [1]{%
 \@ifx{#1\undefined}
}%
\providecommand \@ifnum [1]{%
 \ifnum #1\expandafter \@firstoftwo
 \else \expandafter \@secondoftwo
 \fi
}%
\providecommand \@ifx [1]{%
 \ifx #1\expandafter \@firstoftwo
 \else \expandafter \@secondoftwo
 \fi
}%
\providecommand \natexlab [1]{#1}%
\providecommand \enquote  [1]{``#1''}%
\providecommand \bibnamefont  [1]{#1}%
\providecommand \bibfnamefont [1]{#1}%
\providecommand \citenamefont [1]{#1}%
\providecommand \href@noop [0]{\@secondoftwo}%
\providecommand \href [0]{\begingroup \@sanitize@url \@href}%
\providecommand \@href[1]{\@@startlink{#1}\@@href}%
\providecommand \@@href[1]{\endgroup#1\@@endlink}%
\providecommand \@sanitize@url [0]{\catcode `\\12\catcode `\$12\catcode
  `\&12\catcode `\#12\catcode `\^12\catcode `\_12\catcode `\%12\relax}%
\providecommand \@@startlink[1]{}%
\providecommand \@@endlink[0]{}%
\providecommand \url  [0]{\begingroup\@sanitize@url \@url }%
\providecommand \@url [1]{\endgroup\@href {#1}{\urlprefix }}%
\providecommand \urlprefix  [0]{URL }%
\providecommand \Eprint [0]{\href }%
\providecommand \doibase [0]{http://dx.doi.org/}%
\providecommand \selectlanguage [0]{\@gobble}%
\providecommand \bibinfo  [0]{\@secondoftwo}%
\providecommand \bibfield  [0]{\@secondoftwo}%
\providecommand \translation [1]{[#1]}%
\providecommand \BibitemOpen [0]{}%
\providecommand \bibitemStop [0]{}%
\providecommand \bibitemNoStop [0]{.\EOS\space}%
\providecommand \EOS [0]{\spacefactor3000\relax}%
\providecommand \BibitemShut  [1]{\csname bibitem#1\endcsname}%
\let\auto@bib@innerbib\@empty
\bibitem [{\citenamefont {Luque}\ and\ \citenamefont
  {Mart{\'\i}}(1997)}]{luque1997a}%
  \BibitemOpen
  \bibfield  {author} {\bibinfo {author} {\bibfnamefont {A.}~\bibnamefont
  {Luque}}\ and\ \bibinfo {author} {\bibfnamefont {A.}~\bibnamefont
  {Mart{\'\i}}},\ }\href@noop {} {\bibfield  {journal} {\bibinfo  {journal}
  {Phys. Rev. Lett.}\ }\textbf {\bibinfo {volume} {78}},\ \bibinfo {pages}
  {5014} (\bibinfo {year} {1997})}\BibitemShut {NoStop}%
\bibitem [{\citenamefont {Luque}\ \emph {et~al.}(2012)\citenamefont {Luque},
  \citenamefont {Mart{\'\i}},\ and\ \citenamefont {Stanley}}]{luque2012a}%
  \BibitemOpen
  \bibfield  {author} {\bibinfo {author} {\bibfnamefont {A.}~\bibnamefont
  {Luque}}, \bibinfo {author} {\bibfnamefont {A.}~\bibnamefont {Mart{\'\i}}}, \
  and\ \bibinfo {author} {\bibfnamefont {C.}~\bibnamefont {Stanley}},\
  }\href@noop {} {\bibfield  {journal} {\bibinfo  {journal} {Nat. Photon.}\
  }\textbf {\bibinfo {volume} {6}},\ \bibinfo {pages} {146} (\bibinfo {year}
  {2012})}\BibitemShut {NoStop}%
\bibitem [{\citenamefont {Mart{\'\i}}\ \emph {et~al.}(2008)\citenamefont
  {Mart{\'\i}}, \citenamefont {Marr{\'o}n},\ and\ \citenamefont
  {Luque}}]{marti2008a}%
  \BibitemOpen
  \bibfield  {author} {\bibinfo {author} {\bibfnamefont {A.}~\bibnamefont
  {Mart{\'\i}}}, \bibinfo {author} {\bibfnamefont {D.~F.}\ \bibnamefont
  {Marr{\'o}n}}, \ and\ \bibinfo {author} {\bibfnamefont {A.}~\bibnamefont
  {Luque}},\ }\href@noop {} {\bibfield  {journal} {\bibinfo  {journal} {J.
  Appl. Phys.}\ }\textbf {\bibinfo {volume} {103}},\ \bibinfo {pages} {073706}
  (\bibinfo {year} {2008})}\BibitemShut {NoStop}%
\bibitem [{\citenamefont {Shockley}\ and\ \citenamefont
  {Queisser}(1961)}]{shockley1961a}%
  \BibitemOpen
  \bibfield  {author} {\bibinfo {author} {\bibfnamefont {W.}~\bibnamefont
  {Shockley}}\ and\ \bibinfo {author} {\bibfnamefont {H.~J.}\ \bibnamefont
  {Queisser}},\ }\href@noop {} {\bibfield  {journal} {\bibinfo  {journal} {J.
  Appl. Phys.}\ }\textbf {\bibinfo {volume} {32}},\ \bibinfo {pages} {510}
  (\bibinfo {year} {1961})}\BibitemShut {NoStop}%
\bibitem [{\citenamefont {Hashemi}\ \emph {et~al.}(2014)\citenamefont
  {Hashemi}, \citenamefont {Akbari}, \citenamefont {Huotari},\ and\
  \citenamefont {Hakala}}]{hashemi2014a}%
  \BibitemOpen
  \bibfield  {author} {\bibinfo {author} {\bibfnamefont {J.}~\bibnamefont
  {Hashemi}}, \bibinfo {author} {\bibfnamefont {A.}~\bibnamefont {Akbari}},
  \bibinfo {author} {\bibfnamefont {S.}~\bibnamefont {Huotari}}, \ and\
  \bibinfo {author} {\bibfnamefont {M.}~\bibnamefont {Hakala}},\ }\href@noop {}
  {\bibfield  {journal} {\bibinfo  {journal} {Phys. Rev. B}\ }\textbf {\bibinfo
  {volume} {90}},\ \bibinfo {pages} {075154} (\bibinfo {year}
  {2014})}\BibitemShut {NoStop}%
\bibitem [{\citenamefont {Han}\ \emph {et~al.}(2014)\citenamefont {Han},
  \citenamefont {Zhang},\ and\ \citenamefont {Zeng}}]{han2014a}%
  \BibitemOpen
  \bibfield  {author} {\bibinfo {author} {\bibfnamefont {M.}~\bibnamefont
  {Han}}, \bibinfo {author} {\bibfnamefont {X.}~\bibnamefont {Zhang}}, \ and\
  \bibinfo {author} {\bibfnamefont {Z.}~\bibnamefont {Zeng}},\ }\href@noop {}
  {\bibfield  {journal} {\bibinfo  {journal} {RSC Adv.}\ }\textbf {\bibinfo
  {volume} {4}},\ \bibinfo {pages} {62380} (\bibinfo {year}
  {2014})}\BibitemShut {NoStop}%
\bibitem [{\citenamefont {Chen}\ \emph {et~al.}(2013)\citenamefont {Chen},
  \citenamefont {Qin}, \citenamefont {Chen}, \citenamefont {Yang},
  \citenamefont {Wang},\ and\ \citenamefont {Huang}}]{chen2013a}%
  \BibitemOpen
  \bibfield  {author} {\bibinfo {author} {\bibfnamefont {P.}~\bibnamefont
  {Chen}}, \bibinfo {author} {\bibfnamefont {M.}~\bibnamefont {Qin}}, \bibinfo
  {author} {\bibfnamefont {H.}~\bibnamefont {Chen}}, \bibinfo {author}
  {\bibfnamefont {C.}~\bibnamefont {Yang}}, \bibinfo {author} {\bibfnamefont
  {Y.}~\bibnamefont {Wang}}, \ and\ \bibinfo {author} {\bibfnamefont
  {F.}~\bibnamefont {Huang}},\ }\href@noop {} {\bibfield  {journal} {\bibinfo
  {journal} {Phys. Status Solidi A}\ }\textbf {\bibinfo {volume} {210}},\
  \bibinfo {pages} {1098} (\bibinfo {year} {2013})}\BibitemShut {NoStop}%
\bibitem [{\citenamefont {Yang}\ \emph {et~al.}(2013)\citenamefont {Yang},
  \citenamefont {Qin}, \citenamefont {Wang}, \citenamefont {Wan}, \citenamefont
  {Huang},\ and\ \citenamefont {Lin}}]{yang2013a}%
  \BibitemOpen
  \bibfield  {author} {\bibinfo {author} {\bibfnamefont {C.}~\bibnamefont
  {Yang}}, \bibinfo {author} {\bibfnamefont {M.}~\bibnamefont {Qin}}, \bibinfo
  {author} {\bibfnamefont {Y.}~\bibnamefont {Wang}}, \bibinfo {author}
  {\bibfnamefont {D.}~\bibnamefont {Wan}}, \bibinfo {author} {\bibfnamefont
  {F.}~\bibnamefont {Huang}}, \ and\ \bibinfo {author} {\bibfnamefont
  {J.}~\bibnamefont {Lin}},\ }\href@noop {} {\bibfield  {journal} {\bibinfo
  {journal} {Sci. Rep.}\ }\textbf {\bibinfo {volume} {3}},\ \bibinfo {pages} {1286} (\bibinfo {year}
  {2013})}\BibitemShut {NoStop}%
\bibitem [{\citenamefont {Semin{\'o}vski}\ \emph {et~al.}(2011)\citenamefont
  {Semin{\'o}vski}, \citenamefont {Palacios},\ and\ \citenamefont
  {Wahn{\'o}n}}]{seminovski2011a}%
  \BibitemOpen
  \bibfield  {author} {\bibinfo {author} {\bibfnamefont {Y.}~\bibnamefont
  {Semin{\'o}vski}}, \bibinfo {author} {\bibfnamefont {P.}~\bibnamefont
  {Palacios}}, \ and\ \bibinfo {author} {\bibfnamefont {P.}~\bibnamefont
  {Wahn{\'o}n}},\ }\href@noop {} {\bibfield  {journal} {\bibinfo  {journal}
  {Thin Solid Films}\ }\textbf {\bibinfo {volume} {519}},\ \bibinfo {pages}
  {7517} (\bibinfo {year} {2011})}\BibitemShut {NoStop}%
\bibitem [{\citenamefont {Tablero}\ and\ \citenamefont
  {Fuertes~Marr{\'o}n}(2010)}]{tablero2010a}%
  \BibitemOpen
  \bibfield  {author} {\bibinfo {author} {\bibfnamefont {C.}~\bibnamefont
  {Tablero}}\ and\ \bibinfo {author} {\bibfnamefont {D.}~\bibnamefont
  {Fuertes~Marr{\'o}n}},\ }\href@noop {} {\bibfield  {journal} {\bibinfo
  {journal} {J. Phys. Chem. C}\ }\textbf {\bibinfo {volume} {114}},\ \bibinfo
  {pages} {2756} (\bibinfo {year} {2010})}\BibitemShut {NoStop}%
\bibitem [{\citenamefont {Tablero}(2010)}]{tablero2010b}%
  \BibitemOpen
  \bibfield  {author} {\bibinfo {author} {\bibfnamefont {C.}~\bibnamefont
  {Tablero}},\ }\href@noop {} {\bibfield  {journal} {\bibinfo  {journal} {Thin
  Solid Films}\ }\textbf {\bibinfo {volume} {519}},\ \bibinfo {pages} {1435}
  (\bibinfo {year} {2010})}\BibitemShut {NoStop}%
\bibitem [{\citenamefont {Aguilera}\ \emph {et~al.}(2010)\citenamefont
  {Aguilera}, \citenamefont {Palacios},\ and\ \citenamefont
  {Wahn{\'o}n}}]{aguilera2010a}%
  \BibitemOpen
  \bibfield  {author} {\bibinfo {author} {\bibfnamefont {I.}~\bibnamefont
  {Aguilera}}, \bibinfo {author} {\bibfnamefont {P.}~\bibnamefont {Palacios}},
  \ and\ \bibinfo {author} {\bibfnamefont {P.}~\bibnamefont {Wahn{\'o}n}},\
  }\href@noop {} {\bibfield  {journal} {\bibinfo  {journal} {Solar Energy
  Materials and Solar Cells}\ }\textbf {\bibinfo {volume} {94}},\ \bibinfo
  {pages} {1903} (\bibinfo {year} {2010})}\BibitemShut {NoStop}%
\bibitem [{\citenamefont {Aguilera}\ \emph {et~al.}(2008)\citenamefont
  {Aguilera}, \citenamefont {Palacios},\ and\ \citenamefont
  {Wahn{\'o}n}}]{aguilera2008a}%
  \BibitemOpen
  \bibfield  {author} {\bibinfo {author} {\bibfnamefont {I.}~\bibnamefont
  {Aguilera}}, \bibinfo {author} {\bibfnamefont {P.}~\bibnamefont {Palacios}},
  \ and\ \bibinfo {author} {\bibfnamefont {P.}~\bibnamefont {Wahn{\'o}n}},\
  }\href@noop {} {\bibfield  {journal} {\bibinfo  {journal} {Thin Solid Films}\
  }\textbf {\bibinfo {volume} {516}},\ \bibinfo {pages} {7055} (\bibinfo {year}
  {2008})}\BibitemShut {NoStop}%
\bibitem [{\citenamefont {Palacios}\ \emph {et~al.}(2008)\citenamefont
  {Palacios}, \citenamefont {Aguilera}, \citenamefont {Wahn{\'o}n},\ and\
  \citenamefont {Conesa}}]{palacios2008a}%
  \BibitemOpen
  \bibfield  {author} {\bibinfo {author} {\bibfnamefont {P.}~\bibnamefont
  {Palacios}}, \bibinfo {author} {\bibfnamefont {I.}~\bibnamefont {Aguilera}},
  \bibinfo {author} {\bibfnamefont {P.}~\bibnamefont {Wahn{\'o}n}}, \ and\
  \bibinfo {author} {\bibfnamefont {J.~C.}\ \bibnamefont {Conesa}},\
  }\href@noop {} {\bibfield  {journal} {\bibinfo  {journal} {J. Phys. Chem. C}\
  }\textbf {\bibinfo {volume} {112}},\ \bibinfo {pages} {9525} (\bibinfo {year}
  {2008})}\BibitemShut {NoStop}%
\bibitem [{\citenamefont {Palacios}\ \emph {et~al.}(2007)\citenamefont
  {Palacios}, \citenamefont {S{\'a}nchez}, \citenamefont {Conesa},
  \citenamefont {Fern{\'a}ndez},\ and\ \citenamefont
  {Wahn{\'o}n}}]{palacios2007a}%
  \BibitemOpen
  \bibfield  {author} {\bibinfo {author} {\bibfnamefont {P.}~\bibnamefont
  {Palacios}}, \bibinfo {author} {\bibfnamefont {K.}~\bibnamefont
  {S{\'a}nchez}}, \bibinfo {author} {\bibfnamefont {J.}~\bibnamefont {Conesa}},
  \bibinfo {author} {\bibfnamefont {J.}~\bibnamefont {Fern{\'a}ndez}}, \ and\
  \bibinfo {author} {\bibfnamefont {P.}~\bibnamefont {Wahn{\'o}n}},\
  }\href@noop {} {\bibfield  {journal} {\bibinfo  {journal} {Thin Solid Films}\
  }\textbf {\bibinfo {volume} {515}},\ \bibinfo {pages} {6280} (\bibinfo {year}
  {2007})}\BibitemShut {NoStop}%
\bibitem [{\citenamefont {Palacios}\ \emph {et~al.}(2006)\citenamefont
  {Palacios}, \citenamefont {S{\'a}nchez}, \citenamefont {Conesa},\ and\
  \citenamefont {Wahn{\'o}n}}]{palacios2006a}%
  \BibitemOpen
  \bibfield  {author} {\bibinfo {author} {\bibfnamefont {P.}~\bibnamefont
  {Palacios}}, \bibinfo {author} {\bibfnamefont {K.}~\bibnamefont
  {S{\'a}nchez}}, \bibinfo {author} {\bibfnamefont {J.}~\bibnamefont {Conesa}},
  \ and\ \bibinfo {author} {\bibfnamefont {P.}~\bibnamefont {Wahn{\'o}n}},\
  }\href@noop {} {\bibfield  {journal} {\bibinfo  {journal} {Phys. Status
  Solidi A}\ }\textbf {\bibinfo {volume} {203}},\ \bibinfo {pages} {1395}
  (\bibinfo {year} {2006})}\BibitemShut {NoStop}%
\bibitem [{\citenamefont {Zongyan}\ \emph {et~al.}(2014)\citenamefont
  {Zongyan}, \citenamefont {Dacheng},\ and\ \citenamefont
  {Juan}}]{zongyan2014a}%
  \BibitemOpen
  \bibfield  {author} {\bibinfo {author} {\bibfnamefont {Z.}~\bibnamefont
  {Zongyan}}, \bibinfo {author} {\bibfnamefont {Z.}~\bibnamefont {Dacheng}}, \
  and\ \bibinfo {author} {\bibfnamefont {Y.}~\bibnamefont {Juan}},\ }\href@noop
  {} {\bibfield  {journal} {\bibinfo  {journal} {J. Semicond.}\ }\textbf
  {\bibinfo {volume} {35}},\ \bibinfo {pages} {013002} (\bibinfo {year}
  {2014})}\BibitemShut {NoStop}%
\bibitem [{\citenamefont {Lv}\ \emph {et~al.}(2014)\citenamefont {Lv},
  \citenamefont {Yang}, \citenamefont {Li}, \citenamefont {Li}, \citenamefont
  {Yi}, \citenamefont {Wang}, \citenamefont {Niu},\ and\ \citenamefont
  {Zhong}}]{lv2014a}%
  \BibitemOpen
  \bibfield  {author} {\bibinfo {author} {\bibfnamefont {X.}~\bibnamefont
  {Lv}}, \bibinfo {author} {\bibfnamefont {S.}~\bibnamefont {Yang}}, \bibinfo
  {author} {\bibfnamefont {M.}~\bibnamefont {Li}}, \bibinfo {author}
  {\bibfnamefont {H.}~\bibnamefont {Li}}, \bibinfo {author} {\bibfnamefont
  {J.}~\bibnamefont {Yi}}, \bibinfo {author} {\bibfnamefont {M.}~\bibnamefont
  {Wang}}, \bibinfo {author} {\bibfnamefont {G.}~\bibnamefont {Niu}}, \ and\
  \bibinfo {author} {\bibfnamefont {J.}~\bibnamefont {Zhong}},\ }\href@noop {}
  {\bibfield  {journal} {\bibinfo  {journal} {Solar Energy}\ }\textbf {\bibinfo
  {volume} {103}},\ \bibinfo {pages} {480} (\bibinfo {year}
  {2014})}\BibitemShut {NoStop}%
\bibitem [{\citenamefont {Marsen}\ \emph {et~al.}(2012)\citenamefont {Marsen},
  \citenamefont {Klemz}, \citenamefont {Unold},\ and\ \citenamefont
  {Schock}}]{marsen2012a}%
  \BibitemOpen
  \bibfield  {author} {\bibinfo {author} {\bibfnamefont {B.}~\bibnamefont
  {Marsen}}, \bibinfo {author} {\bibfnamefont {S.}~\bibnamefont {Klemz}},
  \bibinfo {author} {\bibfnamefont {T.}~\bibnamefont {Unold}}, \ and\ \bibinfo
  {author} {\bibfnamefont {H.-W.}\ \bibnamefont {Schock}},\ }\href@noop {}
  {\bibfield  {journal} {\bibinfo  {journal} {Prog. Photovolt: Res. Appl.}\
  }\textbf {\bibinfo {volume} {20}},\ \bibinfo {pages} {625} (\bibinfo {year}
  {2012})}\BibitemShut {NoStop}%
\bibitem [{\citenamefont {Teranishi}\ \emph {et~al.}(1974)\citenamefont
  {Teranishi}, \citenamefont {Sato},\ and\ \citenamefont
  {Kondo}}]{teranishi1974a}%
  \BibitemOpen
  \bibfield  {author} {\bibinfo {author} {\bibfnamefont {T.}~\bibnamefont
  {Teranishi}}, \bibinfo {author} {\bibfnamefont {K.}~\bibnamefont {Sato}}, \
  and\ \bibinfo {author} {\bibfnamefont {K.}~\bibnamefont {Kondo}},\
  }\href@noop {} {\bibfield  {journal} {\bibinfo  {journal} {J. Phys. Soc.
  Jpn.}\ }\textbf {\bibinfo {volume} {36}},\ \bibinfo {pages} {1618} (\bibinfo
  {year} {1974})}\BibitemShut {NoStop}%
\bibitem [{\citenamefont {Von~Bardeleben}\ \emph {et~al.}(1978)\citenamefont
  {Von~Bardeleben}, \citenamefont {Goltzene}, \citenamefont {Meyer},\ and\
  \citenamefont {Schwab}}]{bardeleben1978a}%
  \BibitemOpen
  \bibfield  {author} {\bibinfo {author} {\bibfnamefont {H.}~\bibnamefont
  {Von~Bardeleben}}, \bibinfo {author} {\bibfnamefont {A.}~\bibnamefont
  {Goltzene}}, \bibinfo {author} {\bibfnamefont {B.}~\bibnamefont {Meyer}}, \
  and\ \bibinfo {author} {\bibfnamefont {C.}~\bibnamefont {Schwab}},\
  }\href@noop {} {\bibfield  {journal} {\bibinfo  {journal} {Phys. Status
  Solidi A}\ }\textbf {\bibinfo {volume} {48}},\ \bibinfo {pages} {K145}
  (\bibinfo {year} {1978})}\BibitemShut {NoStop}%
\bibitem [{\citenamefont {Sato}\ and\ \citenamefont
  {Teranishi}(1980)}]{sato1980a}%
  \BibitemOpen
  \bibfield  {author} {\bibinfo {author} {\bibfnamefont {K.}~\bibnamefont
  {Sato}}\ and\ \bibinfo {author} {\bibfnamefont {T.}~\bibnamefont
  {Teranishi}},\ }\href@noop {} {\bibfield  {journal} {\bibinfo  {journal}
  {Jpn. J. Appl. Phys.}\ }\textbf {\bibinfo {volume} {19}},\ \bibinfo {pages}
  {101} (\bibinfo {year} {1980})}\BibitemShut {NoStop}%
\bibitem [{\citenamefont {Tanaka}\ \emph {et~al.}(1989)\citenamefont {Tanaka},
  \citenamefont {Ishii}, \citenamefont {Matsuda}, \citenamefont {Hasegawa},\
  and\ \citenamefont {Sato}}]{tanaka1989a}%
  \BibitemOpen
  \bibfield  {author} {\bibinfo {author} {\bibfnamefont {K.}~\bibnamefont
  {Tanaka}}, \bibinfo {author} {\bibfnamefont {K.}~\bibnamefont {Ishii}},
  \bibinfo {author} {\bibfnamefont {S.}~\bibnamefont {Matsuda}}, \bibinfo
  {author} {\bibfnamefont {Y.}~\bibnamefont {Hasegawa}}, \ and\ \bibinfo
  {author} {\bibfnamefont {K.}~\bibnamefont {Sato}},\ }\href@noop {} {\bibfield
   {journal} {\bibinfo  {journal} {Jpn. J. Appl. Phys.}\ }\textbf {\bibinfo
  {volume} {28}},\ \bibinfo {pages} {12} (\bibinfo {year} {1989})}\BibitemShut
  {NoStop}%
\bibitem [{\citenamefont {Perdew}\ \emph {et~al.}(1996)\citenamefont {Perdew},
  \citenamefont {Burke},\ and\ \citenamefont {Ernzerhof}}]{perdew1996a}%
  \BibitemOpen
  \bibfield  {author} {\bibinfo {author} {\bibfnamefont {J.~P.}\ \bibnamefont
  {Perdew}}, \bibinfo {author} {\bibfnamefont {K.}~\bibnamefont {Burke}}, \
  and\ \bibinfo {author} {\bibfnamefont {M.}~\bibnamefont {Ernzerhof}},\
  }\href@noop {} {\bibfield  {journal} {\bibinfo  {journal} {Phys. Rev. Lett.}\
  }\textbf {\bibinfo {volume} {77}},\ \bibinfo {pages} {3865} (\bibinfo {year}
  {1996})}\BibitemShut {NoStop}%
\bibitem [{\citenamefont {Heyd}\ \emph {et~al.}(2003)\citenamefont {Heyd},
  \citenamefont {Scuseria},\ and\ \citenamefont {Ernzerhof}}]{heyd2003a}%
  \BibitemOpen
  \bibfield  {author} {\bibinfo {author} {\bibfnamefont {J.}~\bibnamefont
  {Heyd}}, \bibinfo {author} {\bibfnamefont {G.~E.}\ \bibnamefont {Scuseria}},
  \ and\ \bibinfo {author} {\bibfnamefont {M.}~\bibnamefont {Ernzerhof}},\
  }\href@noop {} {\bibfield  {journal} {\bibinfo  {journal} {J. Chem. Phys.}\
  }\textbf {\bibinfo {volume} {118}},\ \bibinfo {pages} {8207} (\bibinfo {year}
  {2003})}\BibitemShut {NoStop}%
\bibitem [{\citenamefont {Heyd}\ \emph {et~al.}(2006)\citenamefont {Heyd},
  \citenamefont {Scuseria},\ and\ \citenamefont {Ernzerhof}}]{heyd2006a}%
  \BibitemOpen
  \bibfield  {author} {\bibinfo {author} {\bibfnamefont {J.}~\bibnamefont
  {Heyd}}, \bibinfo {author} {\bibfnamefont {G.~E.}\ \bibnamefont {Scuseria}},
  \ and\ \bibinfo {author} {\bibfnamefont {M.}~\bibnamefont {Ernzerhof}},\
  }\href@noop {} {\bibfield  {journal} {\bibinfo  {journal} {J. Chem. Phys.}\
  }\textbf {\bibinfo {volume} {124}},\ \bibinfo {pages} {219906} (\bibinfo
  {year} {2006})}\BibitemShut {NoStop}%
\bibitem [{\citenamefont {Shay}\ \emph {et~al.}(1972)\citenamefont {Shay},
  \citenamefont {Tell}, \citenamefont {Kasper},\ and\ \citenamefont
  {Schiavone}}]{shay1972a}%
  \BibitemOpen
  \bibfield  {author} {\bibinfo {author} {\bibfnamefont {J.~L.}\ \bibnamefont
  {Shay}}, \bibinfo {author} {\bibfnamefont {B.}~\bibnamefont {Tell}}, \bibinfo
  {author} {\bibfnamefont {H.~M.}\ \bibnamefont {Kasper}}, \ and\ \bibinfo
  {author} {\bibfnamefont {L.~M.}\ \bibnamefont {Schiavone}},\ }\href {\doibase
  10.1103/PhysRevB.5.5003} {\bibfield  {journal} {\bibinfo  {journal} {Phys.
  Rev. B}\ }\textbf {\bibinfo {volume} {5}},\ \bibinfo {pages} {5003} (\bibinfo
  {year} {1972})}\BibitemShut {NoStop}%
\bibitem [{\citenamefont {Tell}\ \emph {et~al.}(1971)\citenamefont {Tell},
  \citenamefont {Shay},\ and\ \citenamefont {Kasper}}]{tell1971a}%
  \BibitemOpen
  \bibfield  {author} {\bibinfo {author} {\bibfnamefont {B.}~\bibnamefont
  {Tell}}, \bibinfo {author} {\bibfnamefont {J.~L.}\ \bibnamefont {Shay}}, \
  and\ \bibinfo {author} {\bibfnamefont {H.~M.}\ \bibnamefont {Kasper}},\
  }\href {\doibase 10.1103/PhysRevB.4.2463} {\bibfield  {journal} {\bibinfo
  {journal} {Phys. Rev. B}\ }\textbf {\bibinfo {volume} {4}},\ \bibinfo {pages}
  {2463} (\bibinfo {year} {1971})}\BibitemShut {NoStop}%
\bibitem [{\citenamefont {Bellabarba}\ \emph {et~al.}(1996)\citenamefont
  {Bellabarba}, \citenamefont {Gonz\'alez},\ and\ \citenamefont
  {Rinc\'on}}]{bellabarba1996a}%
  \BibitemOpen
  \bibfield  {author} {\bibinfo {author} {\bibfnamefont {C.}~\bibnamefont
  {Bellabarba}}, \bibinfo {author} {\bibfnamefont {J.}~\bibnamefont
  {Gonz\'alez}}, \ and\ \bibinfo {author} {\bibfnamefont {C.}~\bibnamefont
  {Rinc\'on}},\ }\href {\doibase 10.1103/PhysRevB.53.7792} {\bibfield
  {journal} {\bibinfo  {journal} {Phys. Rev. B}\ }\textbf {\bibinfo {volume}
  {53}},\ \bibinfo {pages} {7792} (\bibinfo {year} {1996})}\BibitemShut
  {NoStop}%
\bibitem [{\citenamefont {Syrbu}\ \emph {et~al.}(2005)\citenamefont {Syrbu},
  \citenamefont {Tiginyanu}, \citenamefont {Nemerenco}, \citenamefont {Ursaki},
  \citenamefont {Tezlevan},\ and\ \citenamefont {Zalamai}}]{syrbu2005a}%
  \BibitemOpen
  \bibfield  {author} {\bibinfo {author} {\bibfnamefont {N.}~\bibnamefont
  {Syrbu}}, \bibinfo {author} {\bibfnamefont {I.}~\bibnamefont {Tiginyanu}},
  \bibinfo {author} {\bibfnamefont {L.}~\bibnamefont {Nemerenco}}, \bibinfo
  {author} {\bibfnamefont {V.}~\bibnamefont {Ursaki}}, \bibinfo {author}
  {\bibfnamefont {V.}~\bibnamefont {Tezlevan}}, \ and\ \bibinfo {author}
  {\bibfnamefont {V.}~\bibnamefont {Zalamai}},\ }\href@noop {} {\bibfield
  {journal} {\bibinfo  {journal} {J. Phys. Chem. Solids}\ }\textbf {\bibinfo
  {volume} {66}},\ \bibinfo {pages} {1974} (\bibinfo {year}
  {2005})}\BibitemShut {NoStop}%
\bibitem [{\citenamefont {Botha}\ \emph {et~al.}(2007)\citenamefont {Botha},
  \citenamefont {Branch}, \citenamefont {Berndt}, \citenamefont {Leitch},\ and\
  \citenamefont {Weber}}]{botha2007a}%
  \BibitemOpen
  \bibfield  {author} {\bibinfo {author} {\bibfnamefont {J.}~\bibnamefont
  {Botha}}, \bibinfo {author} {\bibfnamefont {M.}~\bibnamefont {Branch}},
  \bibinfo {author} {\bibfnamefont {P.}~\bibnamefont {Berndt}}, \bibinfo
  {author} {\bibfnamefont {A.}~\bibnamefont {Leitch}}, \ and\ \bibinfo {author}
  {\bibfnamefont {J.}~\bibnamefont {Weber}},\ }\href@noop {} {\bibfield
  {journal} {\bibinfo  {journal} {Thin Solid Films}\ }\textbf {\bibinfo
  {volume} {515}},\ \bibinfo {pages} {6246} (\bibinfo {year}
  {2007})}\BibitemShut {NoStop}%
\bibitem [{\citenamefont {Bl{\"o}chl}(1994)}]{blochl1994a}%
  \BibitemOpen
  \bibfield  {author} {\bibinfo {author} {\bibfnamefont {P.~E.}\ \bibnamefont
  {Bl{\"o}chl}},\ }\href@noop {} {\bibfield  {journal} {\bibinfo  {journal}
  {Phys. Rev. B}\ }\textbf {\bibinfo {volume} {50}},\ \bibinfo {pages} {17953}
  (\bibinfo {year} {1994})}\BibitemShut {NoStop}%
\bibitem [{\citenamefont {Kresse}\ and\ \citenamefont
  {Furthm{\"u}ller}(1996)}]{kresse1996a}%
  \BibitemOpen
  \bibfield  {author} {\bibinfo {author} {\bibfnamefont {G.}~\bibnamefont
  {Kresse}}\ and\ \bibinfo {author} {\bibfnamefont {J.}~\bibnamefont
  {Furthm{\"u}ller}},\ }\href@noop {} {\bibfield  {journal} {\bibinfo
  {journal} {Phys. Rev. B}\ }\textbf {\bibinfo {volume} {54}},\ \bibinfo
  {pages} {11169} (\bibinfo {year} {1996})}\BibitemShut {NoStop}%
\bibitem [{\citenamefont {Aguilera}\ \emph {et~al.}(2011)\citenamefont
  {Aguilera}, \citenamefont {Vidal}, \citenamefont {Wahn{\'o}n}, \citenamefont
  {Reining},\ and\ \citenamefont {Botti}}]{aguilera2011a}%
  \BibitemOpen
  \bibfield  {author} {\bibinfo {author} {\bibfnamefont {I.}~\bibnamefont
  {Aguilera}}, \bibinfo {author} {\bibfnamefont {J.}~\bibnamefont {Vidal}},
  \bibinfo {author} {\bibfnamefont {P.}~\bibnamefont {Wahn{\'o}n}}, \bibinfo
  {author} {\bibfnamefont {L.}~\bibnamefont {Reining}}, \ and\ \bibinfo
  {author} {\bibfnamefont {S.}~\bibnamefont {Botti}},\ }\href@noop {}
  {\bibfield  {journal} {\bibinfo  {journal} {Phys. Rev. B}\ }\textbf {\bibinfo
  {volume} {84}},\ \bibinfo {pages} {085145} (\bibinfo {year}
  {2011})}\BibitemShut {NoStop}%
\bibitem [{\citenamefont {Kumakura}\ \emph {et~al.}(2000)\citenamefont
  {Kumakura}, \citenamefont {Iida}, \citenamefont {Nakagaki}, \citenamefont
  {Uchiki}, \citenamefont {Matsumoto-Aoki},\ and\ \citenamefont
  {Kato}}]{kumakura2000a}%
  \BibitemOpen
  \bibfield  {author} {\bibinfo {author} {\bibfnamefont {H.}~\bibnamefont
  {Kumakura}}, \bibinfo {author} {\bibfnamefont {S.}~\bibnamefont {Iida}},
  \bibinfo {author} {\bibfnamefont {Y.}~\bibnamefont {Nakagaki}}, \bibinfo
  {author} {\bibfnamefont {H.}~\bibnamefont {Uchiki}}, \bibinfo {author}
  {\bibfnamefont {T.}~\bibnamefont {Matsumoto-Aoki}}, \ and\ \bibinfo {author}
  {\bibfnamefont {A.}~\bibnamefont {Kato}},\ }\href@noop {} {\bibfield
  {journal} {\bibinfo  {journal} {Jpn. J. Appl. Phys.}\ }\textbf {\bibinfo
  {volume} {39}},\ \bibinfo {pages} {208} (\bibinfo {year} {2000})}\BibitemShut
  {NoStop}%
\end{thebibliography}
%

\end{document}